\documentclass[a4paper,11pt]{article}
\pdfoutput=1 

\usepackage{jcappub} 

\usepackage[T1]{fontenc} 

\title{\boldmath A Variable-Slope Smooth-$k$ Filter for Modeling Halo Abundances with Damped and Oscillatory Power Spectra}

\author{Andreu Rocamora Martorell} 

\affiliation{Facultat de Física, Universitat de Barcelona, Martí Franquès 1, E-08028 Barcelona, Spain}

\emailAdd{andreu.rocamora@ceab.csic.es}

\abstract{We introduce a variable-slope smooth-$k$ (VSMK) filter within the Press-Schechter formalism to model halo mass functions derived from damped and oscillatory matter power spectra. While the standard smooth-$k$ approach successfully captures small-scale suppression effects, it intrinsically couples these to oscillatory features at intermediate scales. The VSMK filter generalizes this framework by allowing the effective logarithmic slope of the $k$-space window function to vary smoothly between two asymptotic regimes, thereby decoupling the small-scale suppression of halo abundances from the intermediate-scale oscillatory features characteristic of dark acoustic oscillations (DAO). We compare the analytic predictions obtained with the VSMK filter to $N$-body simulations for warm dark matter and ETHOS models with DAO, showing that a single parameter set reproduces both regimes simultaneously. The VSMK filter thus provides a unified and flexible analytic framework for modeling halo abundances in non-cold dark matter scenarios with damped and oscillatory power spectra.}

\keywords{dark matter theory, non-cold dark matter, halo mass function, structure formation}
\arxivnumber{2602.01320}

\begin{document}
\maketitle
\flushbottom

\section{Introduction}

Dark matter constitutes the dominant component of the matter density of the Universe, yet its fundamental nature remains one of the central open problems in modern cosmology. The cold dark matter (CDM) paradigm \cite{Peebles1982} has been remarkably successful in describing the abundance of massive cosmic structures, as well as a wide range of large-scale observables \cite{Springel2005,Driver2022}. However, on galactic and sub-galactic scales, CDM faces a number of well-known challenges, including the missing satellites problem \cite{Klypin1999,Moore1999}, the core-cusp problem \cite{Moore1994,deBlok2010}, and the too-big-to-fail problem \cite{BoylanKolchin2011,Papastergis2015}. For comprehensive reviews, see \cite{Weinberg2015,Bullock2017}.

\looseness=-1
Motivated by these small-scale tensions, a broad class of alternative dark matter models has been proposed over the past decades. Many of these scenarios predict a suppression of structure formation on small scales, typically reflected as a cutoff or damping feature in the halo mass function (HMF). In most cases, this suppression originates from physical processes acting after inflation. Examples include free streaming induced by non-negligible thermal velocities, as in warm dark matter (WDM) models \cite{Shi1999,Dodelson1994,Bode2001,Viel2005}; collisional interactions between dark matter and relativistic species, which can give rise to dark acoustic oscillations (DAO) \cite{Sameie2019,Schaeffer2021}; and the quantum wave-like nature of dark matter in fuzzy dark matter (FDM) models \cite{Marsh2016,Schive2016,Hui2017,Dentler2022}.

Current observational constraints on the HMF increasingly push the characteristic suppression scale toward lower halo masses, thereby placing progressively stronger bounds on the minimum dark matter particle mass \cite{Hsueh2019,Shirasaki2021,Sipple2025}. Potentially observable deviations from the CDM prediction are therefore expected to become increasingly important at halo masses $\lesssim 10^7\,M_{\odot}$. Probing this regime requires increasingly sensitive observational techniques, including strong gravitational lensing \cite{Despali2022,WagnerCarena2024,Keeley2024,Vegetti2024,Despali2025}, perturbations of stellar streams \cite{Banik2021,Barry2023,Bonaca2025}, and neutral hydrogen (HI) surveys \cite{Jones2018,Zhang2024,Garland2024}. \\

In practice, numerical simulations are often employed to study the HMF in this mass range. However, simulations face significant challenges at small scales, as low-mass halos can only be reliably identified once they contain a sufficiently large number of particles, giving rise to the well-known issue of spurious halo formation (e.g. \cite{Bose2016,Schive2016,Leo2018}). From a semi-analytical perspective, the extended Press-Schechter (EPS) formalism
\cite{Press1974,Bond1991,Sheth2002} provides a flexible framework to model the HMF in
CDM \cite{Lukic2007,Schneider2012}, WDM \cite{Benson2013,Bose2016}, DAO
\cite{Sameie2019,Schaeffer2021,Verwohlt2024}, and FDM scenarios
\cite{Schneider2018,Kulkarni2021}. The predictive power of the formalism nevertheless
depends on several quantities that are typically calibrated against numerical simulations,
including the collapse model and the relation between halo mass and the density-field filter scale.

The role of density-field filtering in shaping the HMF within the EPS formalism has been extensively studied over the past three decades \cite{Bond1991}. More recently, it has been shown that, for models with a damped linear matter power spectrum, such as WDM or DAO, the small-scale behavior of the HMF is strongly sensitive to the choice of filter \cite{Schaeffer2021,Schneider2013}. In this context, Leo et al.\ introduced the smooth-$k$ space (SMK) filter \cite{Leo2018}, which provides additional freedom to control the small-mass logarithmic slope, hereafter referred to as \textit{slope}, of the halo mass function. The SMK filter has been shown to successfully reproduce WDM halo abundances across a range of cosmological parameters \cite{Parimbelli2021,Schaeffer2021}. 

However, recent studies employing the effective theory of structure formation (ETHOS) framework \cite{CyrRacine2016} have highlighted important limitations of the SMK filter when applied as a universal prescription for both WDM and DAO models \cite{Schaeffer2021,Verwohlt2024}. These limitations arise because the slope of the SMK filter is characterized by a single parameter, which simultaneously controls the suppression of the HMF at small masses and the smoothing of oscillatory features at intermediate scales. As a consequence, the damping characteristic of WDM models and the oscillations induced by DAO become intrinsically coupled, preventing an independent adjustment of the small- and intermediate-mass regimes of the HMF. This motivates the question of whether a simple analytic filter can be constructed to treat these regimes separately.

In this work, we address this problem by exploiting a key property of the EPS formalism: in models with truncated or damped power spectra, the HMF at small halo masses is primarily controlled by the filtering of spatial scales larger than the characteristic collapse scale of the halo (i.e. small wavenumbers), whereas the HMF at intermediate masses --- where DAO effects are most pronounced --- is sensitive to the filter behavior at intermediate and smaller spatial scales. This inversion between the relevant regions of the filter and the corresponding mass scales of the HMF enables the construction of a variable-slope smooth-$k$ (VSMK) filter. By allowing the effective slope of the filter to vary continuously between two asymptotic values, the VSMK filter can independently reproduce both the small-scale suppression of the HMF and the oscillatory features characteristic of DAO.

We therefore propose the VSMK filter as a unified and flexible analytic tool capable of
describing halo mass functions in both WDM and ETHOS-based models with DAO using a
single set of parameters. The proposed formalism provides a unified analytical framework
for modeling distinct classes of non-cold dark matter scenarios while avoiding the need for
model-specific filter prescriptions. As forthcoming observations \cite{Zhang2024,Despali2025,Bonaca2025}
and higher-resolution simulations probe progressively smaller halo mass scales, the VSMK
framework will offer a natural means of testing and refining predictions for the abundance of
low-mass halos.

This paper is organized as follows. In Section~\ref{sec:PSF}, we summarize the Press-Schechter formalism and the transfer functions used to construct non-CDM power spectra. The variable-slope smooth-$k$ filter is introduced and motivated in Section~\ref{sec:filter}. Our results and comparisons with numerical simulations are presented in Section~\ref{sec:results}, followed by the conclusions in Section~\ref{sec:conclusions}.

\section{The Press-Schechter formalism}\label{sec:PSF}
The Press-Schechter formalism (PSF) \cite{Press1974,Coles1995}, and in particular its extended formulation (EPS) \cite{Bond1991,Sheth2002,Zentner2007,Mo2010,Maggiore2010}, provides an analytic framework to describe several aspects of the non-linear evolution of cosmic structure. The formalism is based on the linear density contrast field $\delta(\mathbf{x},t)$ and assumes that density perturbations grow linearly in time according to
\begin{equation}
    \delta(\mathbf{x},t) = \delta_0(\mathbf{x})\,D(t),
\end{equation}
where $D(t)$ is the linear growth factor, normalized to unity at the present time.

Within this framework, the fraction of mass collapsed into halos of mass $M$ at redshift $z$ is identified with the fraction of trajectories of the smoothed density field that exceed a critical threshold for collapse, $\delta_c(z)$, when filtered on the mass scale $M$ \cite{Press1974}. This assumption leads to an expression for the halo mass function (HMF) given by
\begin{equation}
\label{eq:HMF}
    \frac{dn}{d\ln M} = -\frac{1}{2}\,\frac{\bar{\rho}}{\sigma^2(M)}\,f(\nu)\,
    \frac{d\sigma^2(M)}{dM},
\end{equation}
where $n$ denotes the comoving number density of halos, $\bar{\rho}$ is the mean matter density of the Universe, $\sigma^2(M)$ is the variance of the density field in the scale $M$, and $f(\nu)$ is the first-crossing distribution.

Following previous work in truncated and oscillatory scenarios \cite{Bohr2021,Schaeffer2021,Verwohlt2024}, we assume ellipsoidal collapse \cite{Sheth1999}, for which the first-crossing distribution is well described by
\begin{equation}
\label{eq:first_crossing_distribution}
    f(\nu) = A \sqrt{\frac{2q\nu}{\pi}}
    \left[1 + (q\nu)^{-p}\right]\exp\!\left(-\frac{q\nu}{2}\right),
\end{equation}
with fiducial parameter values $A = 0.322$, $q = 0.707$, and $p = 0.3$. When comparing the analytical model with the simulations of \cite{Bohr2021} (Figure~\ref{fig:results2}), we adopt $A = 0.3658$, consistent with the calibration reported in that work.

The dimensionless peak-height parameter $\nu$ is defined in terms of the critical overdensity for spherical collapse as
\begin{equation}
\label{eq:nu_definition}
    \nu = \frac{\delta_c^2(0)}{\sigma^2(R)\,D^2(z)},
\end{equation}
where $\delta_c(0)$ is the present-day linear collapse threshold, $\sigma^2(R)$ is the variance of the density field on the scale $R$, and $D(z)$ is the linear growth factor evaluated at redshift $z$.

The variance of the density perturbations on the scale $R$ is given by
\begin{equation}
\label{eq:variance}
    \sigma^2(R) = \frac{1}{2\pi^2}
    \int_0^{\infty} k^2\,P(k)\,W^2(k,R)\,dk,
\end{equation}
where $k$ denotes the wavenumber, $P(k)$ is the linear matter power spectrum of the considered dark matter model, and $W(k,R)$ is a spherically symmetric Fourier-space window function for the scale $R$.

\subsection{Density-field filters}
\label{sec:filters}

Window functions, also known as filters, define how density perturbations are weighted as a function of distance from a given point in order to determine whether a halo of mass $M$ collapses. For a halo with a given characteristic scale $R$ and mass $M$, the filtering procedure suppresses the contribution of perturbations on scales smaller than the halo size. In Fourier space, this corresponds to weighting the modes that compose the density contrast field, such that
\begin{equation}
    \delta(k,R) = \hat{\delta}(k)\,W(k,R),
\end{equation}
where $\hat{\delta}(k)$ is the Fourier transform of $\delta(x)$, and $W(k,R)$ is a window function associated with the scale $R$.

The simplest choice of window function is the real-space top-hat (TH) filter,
\begin{equation}
    W_{\rm TH}(r,R) = \Theta(1 - r/R),
\end{equation}
whose main advantage lies in its straightforward physical interpretation. The real-space TH filter has been widely employed in modelling the CDM halo mass function and provides accurate results when calibrated against numerical simulations, e.g. \cite{Tinker2008,Delos2024}. However, in models with a damped linear power spectrum, it produces an unphysical upturn of the HMF at low
masses that is not observed in numerical simulations \cite{Schneider2013}.

A simple alternative that improves the agreement with simulations is the sharp-$k$ space (SHK) filter introduced in \cite{Bond1991},
\begin{equation}
    W_{\rm SHK}(k,R) = \Theta(1 - kR).
\end{equation}
By imposing a sharp cutoff in Fourier space, this filter successfully reproduces the suppression of the HMF in warm dark matter universes \cite{Schneider2013,Bose2016}. Nevertheless, while the SHK filter induces a decline of the HMF at small masses, the slope of this decline is fixed and cannot be tuned.

To allow for greater flexibility in the small-scale behavior of the HMF, the smooth-$k$ space (SMK) filter was proposed in \cite{Leo2018},
\begin{equation}
\label{eq:SMK}
    W_{\rm SMK}(k,k_M^{-1}) = \left[1 + \left(\frac{k}{k_M}\right)^{\beta}\right]^{-1},
\end{equation}
where $k_M = 1/R$. In models with a damped power spectrum, the SMK filter yields a halo mass function that asymptotically follows
\begin{equation}
    \frac{dn}{d\ln M} \propto M^{(\beta - 3)/3}
\end{equation}
at small masses (see Section~\ref{sec:filter}). 

\begin{figure}[htbp] 
    \centering
    
    \begin{minipage}{0.47\textwidth}
        \centering
        \includegraphics[width=\textwidth]{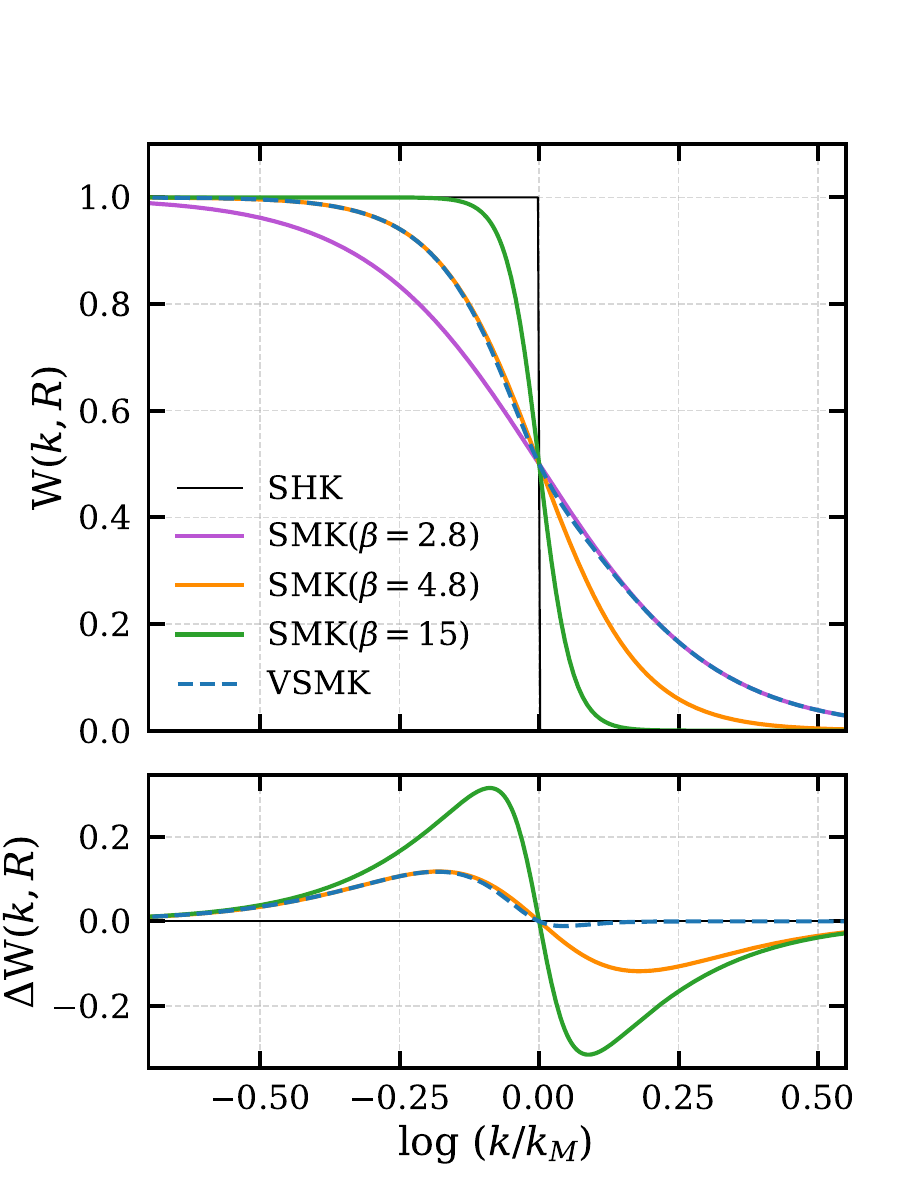}
        \label{fig:sub1}
    \end{minipage}
    
    \caption{\textbf{Comparison between smooth-$k$ (SMK) and variable-slope smooth-$k$ (VSMK) filters.}
    The SMK filter is shown for different values of the slope parameter $\beta$ and compared to a VSMK filter with $\beta_1 = 4.8$, $\beta_2 = 2.8$, $\delta = 12$, and $\mu = 1$, chosen for illustrative purposes. In all cases, $c = 1$ is also adopted for visualization only. The lower panel shows the difference $\Delta W(k,R) = W(k,R) - W_{\mathrm{SMK}(\beta=2.8)}(k,R)$ for each filter.}
    \label{fig:filtres}
\end{figure}

Motivated by this behavior, and in order to disentangle the contributions of different mass regimes in the HMF, we propose a filter that interpolates between two SMK-like behaviors: an effective slope $\beta_1$ for $k/k_M \ll 1$ and a slope $\beta_2$ for $k/k_M \gg 1$. The explicit functional form of this variable-slope smooth-$k$ (VSMK) filter, together with its derivation and properties, is presented in Section~\ref{sec:filter} (see also Figure~\ref{fig:filtres}).

The definition of halo mass is not straightforward for filters such as SHK and SMK, since their real-space counterparts have divergent integrals. To circumvent this issue, it is customary to assume that the mass scale obeys $M \propto R^3$ owing to the spherical symmetry of the filtering procedure, and to define
\begin{equation}
  \label{eq:mass}
    M = \frac{4\pi}{3}\,\bar{\rho}\,(cR)^3,
\end{equation}
where $c$ is a free parameter calibrated against numerical simulations. This parameter has been shown to improve the agreement between analytic predictions and simulations for the HMF \cite{Leo2018,Sameie2019,Schaeffer2021,Verwohlt2024}, halo concentrations \cite{Brown2021} --- using procedures such as those described in \cite{Diemer2015,Ludlow2016} --- and the slope of the Einasto density profile \cite{Brown2021}. 

It is important to note that the PSF does not uniquely specify the operational definition of a dark matter halo. Consequently, quantities calibrated against numerical
simulations, including the mass assignment relation (Eq.~\ref{eq:mass}),
may depend on the halo finder and halo definition adopted in a given analysis,
particularly in WDM and DAO scenarios \cite{Stucker2022}.

Varying $c$ effectively shifts the halo mass function along the mass axis, without altering its asymptotic slopes or its sensitivity to specific features of the linear power spectrum. In this sense, $c$ plays the same role in the VSMK filter as in the standard SMK case, and should be regarded as a mass calibration parameter rather than a shape parameter of the filter. Formally, this follows from the fact that the parameter $c$ does not enter the window function $W(k,R)$ itself, but only the relation between halo mass and the characteristic wavenumber, $k_M = R^{-1}$. As a result, varying $c$ changes the value of $k_M$ at which the variance is evaluated, but does not modify the relative weighting of Fourier modes within the variance integral. This behavior is therefore qualitatively different from that of the slope parameters $\beta_1$ and $\beta_2$ (Section \ref{sec:filter}), which explicitly alter the functional form of the filter and control how oscillatory features of the power spectrum are smoothed.

Originally, a value $c \simeq 2.42$ was proposed for the SHK filter \cite{Lacey1993}. Subsequent studies found that values in the range $c = 2.5$--$2.7$ provide a better fit for SHK-based models \cite{Bose2016}. More recently, analyses employing the SMK filter have identified a preferred range $c = 3$--$3.8$ \cite{Leo2018,Sameie2019,Bohr2021,Schaeffer2021,Verwohlt2024}. In this work, we adopt $c=3.8$ unless stated otherwise (Section~\ref{sec:results}).

\subsection{Linear power spectra in non-CDM models}
\label{sec:power_spectrum_nonCDM}

The linear matter power spectrum of a non-cold dark matter model can, in general, be expressed in terms of a transfer function $T(k)$ as
\begin{equation}
\label{eq:transfer_general}
    P_i(k) = P_{\rm CDM}(k)\,T^2(k),
\end{equation}
where $P_{\rm CDM}(k)$ is the linear CDM power spectrum and $P_i(k)$ denotes the corresponding non-CDM spectrum. The CDM power spectrum used in this work is computed with the \textsc{CAMB} code \cite{Lewis2000}.

A unified description of a wide class of transfer functions was proposed in \cite{Murgia2017} and subsequently adopted as the basis for the ETHOS framework \cite{Bohr2020}. The ETHOS transfer function \cite{CyrRacine2016,Archidiacono2017}, as derived in \cite{Bohr2020}, allows one to describe warm dark matter–like suppression as well as the presence of dark acoustic oscillations (DAO) in a single parametrized form \cite{Bohr2020,Bohr2021,Verwohlt2024}.

In this work, we model the linear power spectrum of WDM using the standard transfer function introduced in \cite{Bode2001},
\begin{equation}
\label{eq:T_WDM}
    T_{\rm WDM}(k) = \left[1 + (\alpha k)^{2\nu}\right]^{-5/\nu},
\end{equation}
with $\nu = 1.12$ \cite{Viel2005}. The parameter
\begin{equation}
\label{eq:alpha_WDM}
    \alpha = 0.049
    \left(\frac{m_{\mathrm{WDM}}}{1~\mathrm{keV}}\right)^{-1.11}
    \left(\frac{\Omega_m}{0.25}\right)^{0.11}
    \left(\frac{h}{0.7}\right)^{1.22}
\end{equation}
sets the characteristic scale of the suppression, while $\nu$ controls its shape.

For models featuring dark acoustic oscillations, we adopt the ETHOS transfer function presented in \cite{Bohr2020},
\begin{multline}
\label{eq:T_DAO}
    T_{\rm ETHOS}(k) =
    \left[ 1 + (\alpha k)^{\beta} \right]^{\gamma}
    - \sqrt{h_{\mathrm{peak}}}\,
    \exp\!\left[-\frac{1}{2}\left(\frac{x - 1}{\Sigma}\right)^2\right]
    + {} \\
    \frac{\sqrt{h_2}}{4}\,
    \operatorname{erfc}\!\left(\frac{x - x_0}{\tau} - 2\right)
    \operatorname{erfc}\!\left(-\frac{x - x_0}{\Sigma} - 2\right)
    \cos\!\left(1.1083\pi x\right),
\end{multline}
where $x \equiv k/k_{\mathrm{peak}}$, $x_0 = 1.805$, and $\Sigma = 0.2$. The parameter $\alpha$ plays a role analogous to the suppression scale in Eq.~\eqref{eq:T_WDM}, while $\beta$ and $\gamma$ determine the shape of the suppression. The parameters $h_{\mathrm{peak}}$ and $k_{\mathrm{peak}}$ control the amplitude and characteristic scale of the first DAO peak, respectively, and $h_2$ sets the amplitude of the second oscillation.

Within this generalized formalism, all parameters are uniquely determined by the pair $(h_{\mathrm{peak}}, k_{\mathrm{peak}})$. In particular, the limit $h_{\mathrm{peak}} = 0$ recovers the WDM-like transfer function in Eq.~\eqref{eq:T_WDM}, while the limit $k_{\mathrm{peak}} \to \infty$ yields $T(k) \to 1$, corresponding to the CDM power spectrum.

Following \cite{Bohr2020}, the parameter $\alpha$ can be expressed as
\begin{equation}
\label{eq:alpha_ETHOS}
    \alpha =
    \frac{d}{k_{\mathrm{peak}}}
    \left[
    \left(\frac{1}{\sqrt{2}}\right)^{1/\gamma}
    - 1
    \right]^{1/\beta},
\end{equation}
where the parameters $d$ and $\beta$ depend on $h_{\mathrm{peak}}$, and $\gamma = -20$. Explicit values for representative cases are provided in Table~\ref{tab:T_Bohr2020}, following \cite{Bohr2020} and the interpolation scheme of \cite{Verwohlt2024}.

Finally, within the ETHOS framework, the WDM particle mass can be related to the ETHOS peak scale through \cite{Viel2005,Bohr2020}
\begin{equation}
\label{eq:mwdm_ethos}
    \frac{m_{\mathrm{WDM}}}{1~\mathrm{keV}} =
    \left[
    0.050
    \left(\frac{k_{\mathrm{peak}}}{h~\mathrm{Mpc}^{-1}}\right)
    \left(\frac{\Omega_{\chi}}{0.25}\right)^{0.11}
    \left(\frac{h}{0.7}\right)^{1.22}
    \right]^{1/1.11}.
\end{equation}

\begin{table}[h]
\centering
\caption{Parameters used to construct the linear matter power spectra within the ETHOS framework, as a function of the amplitude $h_{\mathrm{peak}}$ of the first dark acoustic oscillation. Only the representative cases $h_{\mathrm{peak}} = 0.0$, $0.4$, and $1.0$ from \cite{Bohr2020} are shown.}
\vspace{2mm}
\begin{tabular}{c c c c c c}
\hline
$h_{\text{peak}}$ & $h_2$ & $\tau$ & $\Sigma$ & $\beta$ & $d$ \\
\hline
0.0 & 0.0 & 0.0 & 0.0 & 2.24 & 3.0 \\
0.4 & $0.221 \text{exp}(-0.025k_{peak}) + 0.0129$ & 0.34 & 0.22 & 3.61 & 2.61 \\
1.0 & 1.08 & 0.67 & 0.2 & 4.05 & 2.5 \\
\hline
\end{tabular}
\label{tab:T_Bohr2020}
\end{table}

\section{The variable slope smooth-$k$ filter}
\label{sec:filter}
The standard SMK filter does not allow for an independent control of the low-mass and high-mass regimes of the HMF. As a consequence, it cannot simultaneously reproduce both the small-scale suppression characteristic of models with a damped matter power spectrum and the oscillatory features induced by dark acoustic oscillations (see Section~\ref{sec:results}). 

To construct a more general filter capable of capturing both effects, we exploit a fundamental property of the Press-Schechter formalism: for models with a damped linear power spectrum, the asymptotic slope of the HMF at small masses is determined by the behavior of the filter in the regime $k/k_M \ll 1$. At large masses, the HMF is largely insensitive to the filter and mainly set by the shape of the power spectrum. However, in models exhibiting oscillatory features, such as DAO scenarios, the HMF can retain a residual dependence on the filter around intermediate wavenumbers, typically $k/k_M \gtrsim 1$. These observations, discussed in detail in Appendix~\ref{app:aprox_filter}, motivate the introduction of a filter with independently tunable slopes in these two regimes.

\begin{figure}[htbp]
    \centering
    \begin{minipage}{0.47\textwidth}
        \centering
        \includegraphics[width=\textwidth]{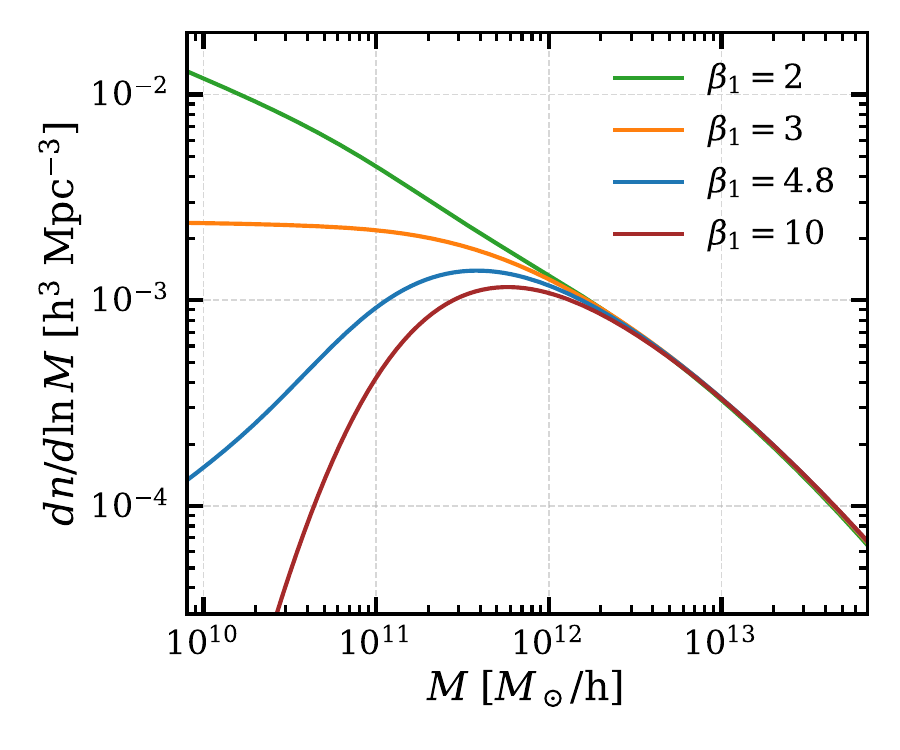}
    \end{minipage}
    \hspace{3mm}
    \begin{minipage}{0.47\textwidth}
        \centering
        \includegraphics[width=\textwidth]{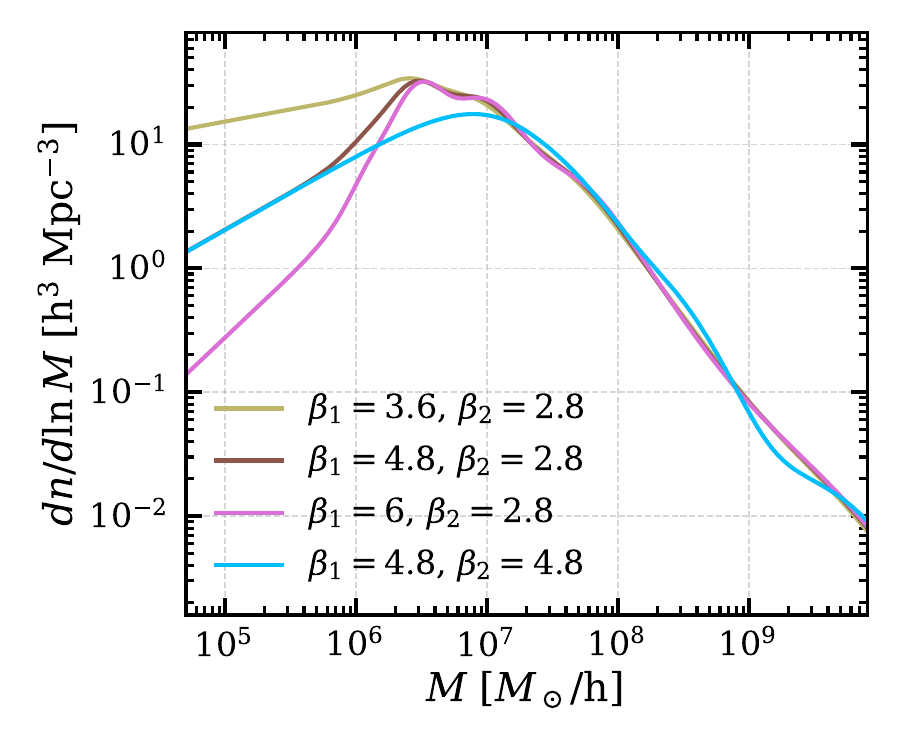}
    \end{minipage}
    \caption{\textbf{Impact of the VSMK filter parameters on the halo mass function.}
    Both panels show halo mass functions computed using the variable-slope smooth-$k$ (VSMK) filter with fixed $\mu = 2.1$, $\delta = 12$, and $c = 3.8$.
    \textit{Left:} WDM-like power spectrum with fixed $\beta_2 = 2.8$ and varying $\beta_1$, illustrating that $\beta_1$ exclusively controls the small-mass slope of the HMF.
    \textit{Right:} DAO power spectrum with simultaneous variations of $\beta_1$ and $\beta_2$, demonstrating that $\beta_1$ regulates the small-mass behavior, while $\beta_2$ controls the intermediate-mass regime where DAO-induced oscillations are present.}
    \label{fig:HMF}
\end{figure}

For sufficiently small masses in models with a damped power spectrum (i.e. WDM or DAO), the variance of the density field asymptotically approaches a constant value \cite{Schneider2013,Bose2016}. In this regime, one finds $\mathrm{d}n/\mathrm{d}\ln M \propto \mathrm{d}\sigma^2/\mathrm{d}k_M$ \cite{Schneider2013,Leo2018}. The derivative of the variance with respect to $k_M$ is given by
\begin{equation}
\label{eq:deriv_var}
    \frac{\mathrm{d} \sigma^2(k_M^{-1})}{\mathrm{d}k_M} =
    \frac{1}{2\pi^2}
    \int_0^{\infty} \xi(k,k_M) \,
    \mathrm{d}k,
\end{equation}
where the integrand is
\begin{equation}
    \label{eq:integrand}
    \xi(k,k_M) \equiv k^2 P(k)\,
    \frac{\partial W^2(k,k_M^{-1})}{\partial k_M} = 2 k^2 P(k)\, W(k,k_M^{-1})
    \frac{\partial W(k,k_M^{-1})}{\partial k_M}.
\end{equation}
For the SMK filter, the integrand $\xi(k,k_M)$ exhibits a pronounced maximum in the regime $k/k_M \ll 1$ when $k_M \gg k_{\mathrm{hm}}$, where $k_{\mathrm{hm}}$ denotes the half-mode scale at which the HMF is suppressed by a factor of two relative to CDM. As the mass decreases, this maximum becomes increasingly localized, allowing the integral in Eq.~\eqref{eq:deriv_var} to be approximated around its peak (Appendix~\ref{app:small_k_small_mass}). Since the VSMK filter asymptotically reduces to the SMK form in this regime (see Figure~\ref{fig:filtres}), the same argument applies to both filters.

Expanding the SMK filter, $W_{SMK} \equiv W_{SMK}(k,R)$, at leading non-vanishing order for $k/k_M \ll 1$, one obtains
\begin{equation}
    W_{\mathrm{SMK}} = 1 - \left( \frac{k}{k_M} \right)^{\beta}
    + \mathcal{O}\!\left[(k/k_M)^{2\beta}\right],
\end{equation}
with the analogous expansion for the VSMK filter (Eqs. ~\eqref{eq:VSMK} and \eqref{eq:variable_beta_VSMK}),
\begin{equation}
    W_{\mathrm{VSMK}} = 1 - \left( \frac{k}{k_M} \right)^{\beta_1}
    + \mathcal{O}\!\left[(k/k_M)^{2\beta_1}\right].
\end{equation}
Substituting these expressions into Eqs.~\eqref{eq:deriv_var} and \eqref{eq:integrand} yields
$\mathrm{d}\sigma^2/\mathrm{d}k_M \propto k_M^{-(\beta_1+1)}$, where $\beta_1$ is replaced by $\beta$ for the SMK filter. Therefore,
$\sigma^2(R) = \sigma_0^2 + C k_M^{-\beta_1}$, consistently recovering the constant-variance limit at small scales. Using the scaling $M \propto k_M^{-3}$ from Eq.~\eqref{eq:mass}, the asymptotic behavior of the HMF becomes
\begin{equation}
\label{eq:small_scale_HMF_trend}
    \frac{\mathrm{d}n}{\mathrm{d}\ln M}
    \propto k_M^{3-\beta_1}
    \propto M^{(\beta_1-3)/3},
\end{equation}
in agreement with previous results for the SMK filter \cite{Schaeffer2021}. This expression is valid up to $\beta_1 \lesssim 22 - n_s$ for power spectra obtained with the transfer function of Viel et al. \eqref{eq:T_WDM}. For larger values of $\beta_1$, the asymptotic behavior at large $k_M$ converges to that of the SHK filter,
$\mathrm{d}n/\mathrm{d}\ln M \propto k_M^6 P_{\mathrm{WDM}}(k_M)$
\cite{Schneider2013} (Appendix \ref{app:small_k_small_mass}).

At larger masses ($k_M < k_{\mathrm{hm}}$), the maximum of $\xi(k,k_M)$ shifts toward $k/k_M \sim 1$, and the previous approximation no longer applies. In this regime, the HMF is largely insensitive to the filter and primarily determined by the shape of the linear power spectrum (Appendix~\ref{sec:appendix_WDM_filter_independence}). However, when the power spectrum exhibits oscillatory features, the filter plays a crucial role in smoothing these oscillations, thereby regulating their imprint on the HMF \cite{Schaeffer2021}. Consequently, the HMF acquires a dependence on the filter around intermediate wavenumbers, $k/k_M \gtrsim 1$ (Appendix~\ref{sec:appendix_DAO_filter_dependence}).

These considerations motivate the definition of the variable-slope smooth-$k$ (VSMK) filter,
\begin{equation}
    \label{eq:VSMK}
    W_{\mathrm{VSMK}} =
    \left[
    1 + \left( \frac{k}{k_M} \right)^{f(k)}
    \right]^{-1},
\end{equation}
where the effective slope $f(k)$ interpolates between two asymptotic values,
\begin{equation}
    \label{eq:variable_beta_VSMK}
    f(k) =
    \beta_2 -
    (\beta_2 - \beta_1)
    \left[
    1 + \left( \mu \frac{k}{k_M} \right)^{\delta}
    \right]^{-1}.
\end{equation}
In this parametrization, $\beta_1$ governs the small-scale suppression of the HMF, while $\beta_2$ controls the intermediate-mass regime where DAO-induced oscillations are relevant. The parameter $\mu$ sets the characteristic transition scale in units of $k_M$, and $\delta$ determines the sharpness of the transition. Larger values of $\delta$ correspond to a more abrupt change of slope, whereas smaller values produce a smoother interpolation. Accordingly, $\mu$ and $\delta$ do not affect the asymptotic behavior of either regime and are therefore treated as fitting parameters controlling the transition only. In this way, this construction introduces the minimal number of additional parameters required to independently control both regimes. As illustrated in Figures~\ref{fig:filtres} and \ref{fig:HMF}, it allows the two asymptotic regimes of the filter to map directly onto two distinct regimes of the HMF.

\section{Comparison with $N$-body simulations and existing SMK calibrations}
\label{sec:results}

To assess the ability of the VSMK filter to reproduce previous SMK descriptions of halo abundances, we consider a set of WDM and DAO models for which calibrated SMK parameterizations have been reported in the literature \cite{Schaeffer2021,Bohr2021,Verwohlt2024}. These calibrations were originally obtained by comparison with $N$-body simulations and therefore provide a useful benchmark against which to evaluate the VSMK formalism. The $N$-body simulations adopt cosmological parameters consistent with \textit{Planck} 2018 results \cite{Planck2020}. In particular, the parameter values $\Omega_m = 0.321$, $\Omega_\Lambda = 0.679$, $h = 0.6688$, $n_s = 0.96$, and $\sigma_8 = 0.811$ are used when comparing analytic predictions with the simulations of \cite{Schaeffer2021,Verwohlt2024}, while the simulations presented in \cite{Bohr2021} employ $\Omega_m = 0.311$, $\Omega_\Lambda = 0.689$, $h = 0.675$, $n_s = 0.965$, and $\sigma_8 = 0.815$.

To construct a representative VSMK calibration, we define one reference SMK parameterization for WDM models and another for DAO models. For the WDM model of \cite{Schaeffer2021}, we adopt the calibration reported in that work, characterized by $\beta = 4.8$ and $c = 3.3$, which corresponds to the original fit from Leo et al. \cite{Leo2018}. For DAO models, we instead determine a representative SMK parameterization, characterized by $\beta = 2.8$ and $c = 3.8$, which provides a good simultaneous description of the simulation data from \cite{Bohr2021,Verwohlt2024} while remaining consistent with the SMK-based descriptions previously reported in these works. With these reference parameterizations established, we can now assess the extent to which a single VSMK prescription is able to reproduce both WDM and DAO halo abundances. \\

\begin{figure}[htbp]
    \centering
    \begin{minipage}{0.49\textwidth}
        \centering
        \includegraphics[width=\textwidth]{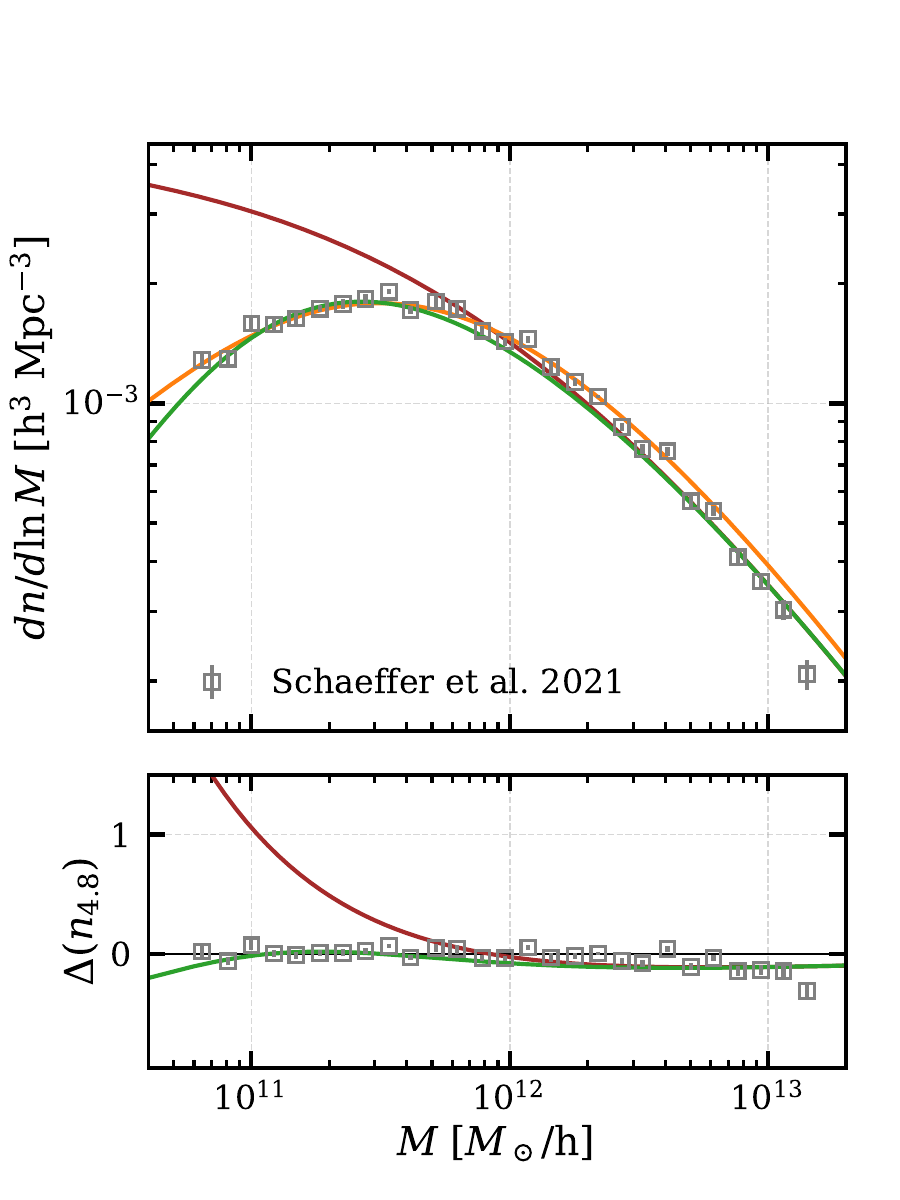}
    \end{minipage}
    \hfill
    \begin{minipage}{0.49\textwidth}
        \centering
        \includegraphics[width=\textwidth]{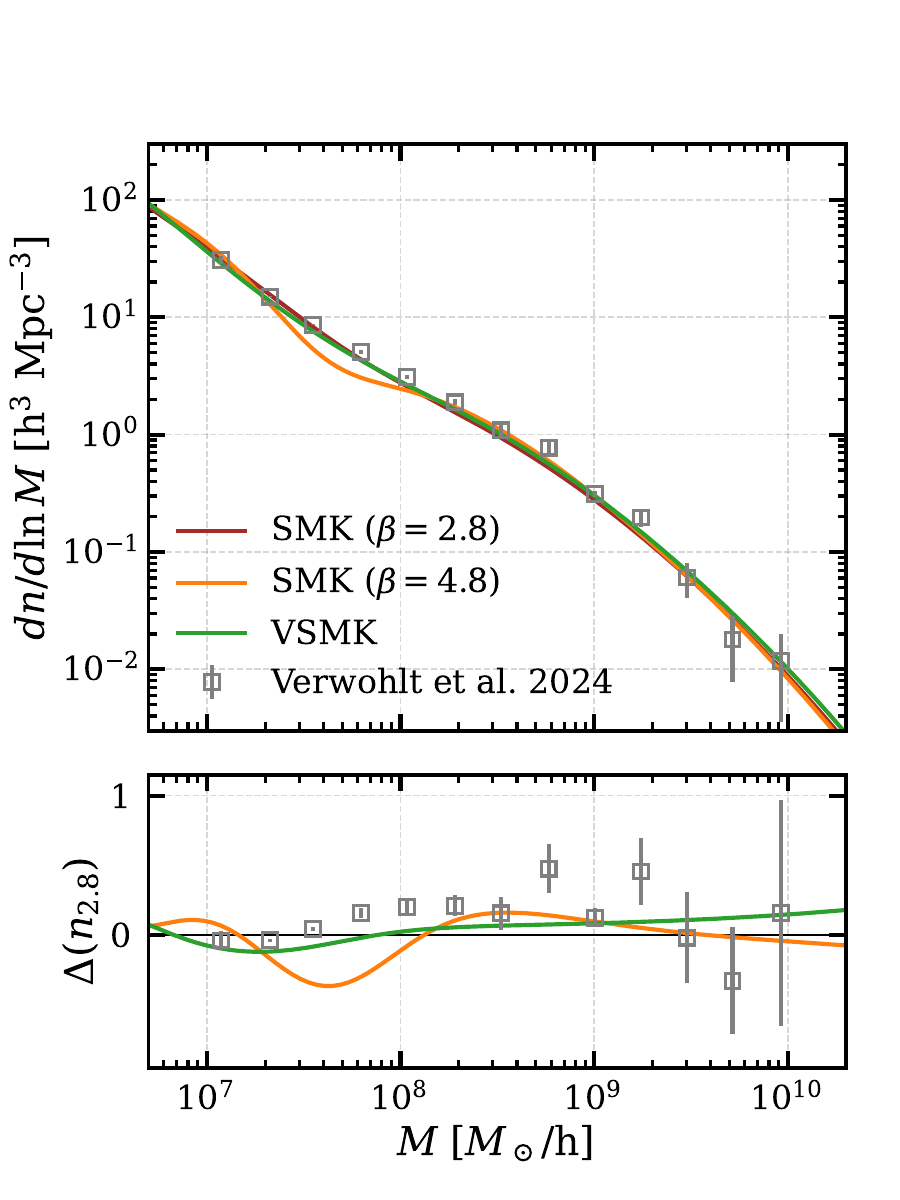}
    \end{minipage}
    \caption{\textbf{Comparison between SMK and VSMK predictions for WDM and DAO models.}
    The halo mass functions obtained with the VSMK filter are compared with those derived from the reference SMK parameterizations adopted for each model. In particular, the HMFs computed using the optimal SMK filter for a given model are contrasted with those obtained using the SMK calibration of the alternative model and with the VSMK prescription. Unless stated otherwise, $c = 3.8$.
    \textit{Left:} WDM model with $m_{\mathrm{WDM}} = 0.25~\mathrm{keV}$ from \cite{Schaeffer2021}, where the optimal SMK filter has $\beta = 4.8$ and $c = 3.3$ (grey dashed line).
    \textit{Right:} DAO model with $h_{\mathrm{peak}} = 1$, $k_{\mathrm{peak}} = 100~h\,\mathrm{Mpc}^{-1}$ at $z=10$ from \cite{Verwohlt2024}, where the optimal SMK filter has $\beta = 2.8$ and $c = 3.8$ (red line).
    Simulation data (empty squares) are extracted from published figures in \cite{Schaeffer2021} and \cite{Verwohlt2024}.
    The lower panels show the relative deviations $\Delta$ with respect to the optimal SMK prediction for each model.}
    \label{fig:results1}
\end{figure}

In both \cite{Schaeffer2021} and \cite{Verwohlt2024} it has been shown that a single SMK filter cannot simultaneously reproduce both the small-scale suppression characteristic of WDM models and the oscillatory features induced by dark acoustic oscillations. Figure~\ref{fig:results1} illustrates this limitation and demonstrates that the VSMK filter is able to describe both scenarios using a single set of parameters, namely $\beta_1 = 4.8$, $\beta_2 = 2.8$, $\mu = 2.1$, $\delta = 12$, and $c = 3.8$. The parameters $\mu$ and $\delta$ are kept fixed across all models and redshifts considered in this work, and are not re-optimized for individual simulations. This choice is motivated by the fact that variations in $\beta_1$ and $\beta_2$ account for the dominant changes in the halo mass function over the range of models considered here. 

\vspace{-1mm}

\begin{figure}[htbp]
    \centering
    \begin{minipage}{0.49\textwidth}
        \centering
        \includegraphics[width=\textwidth]{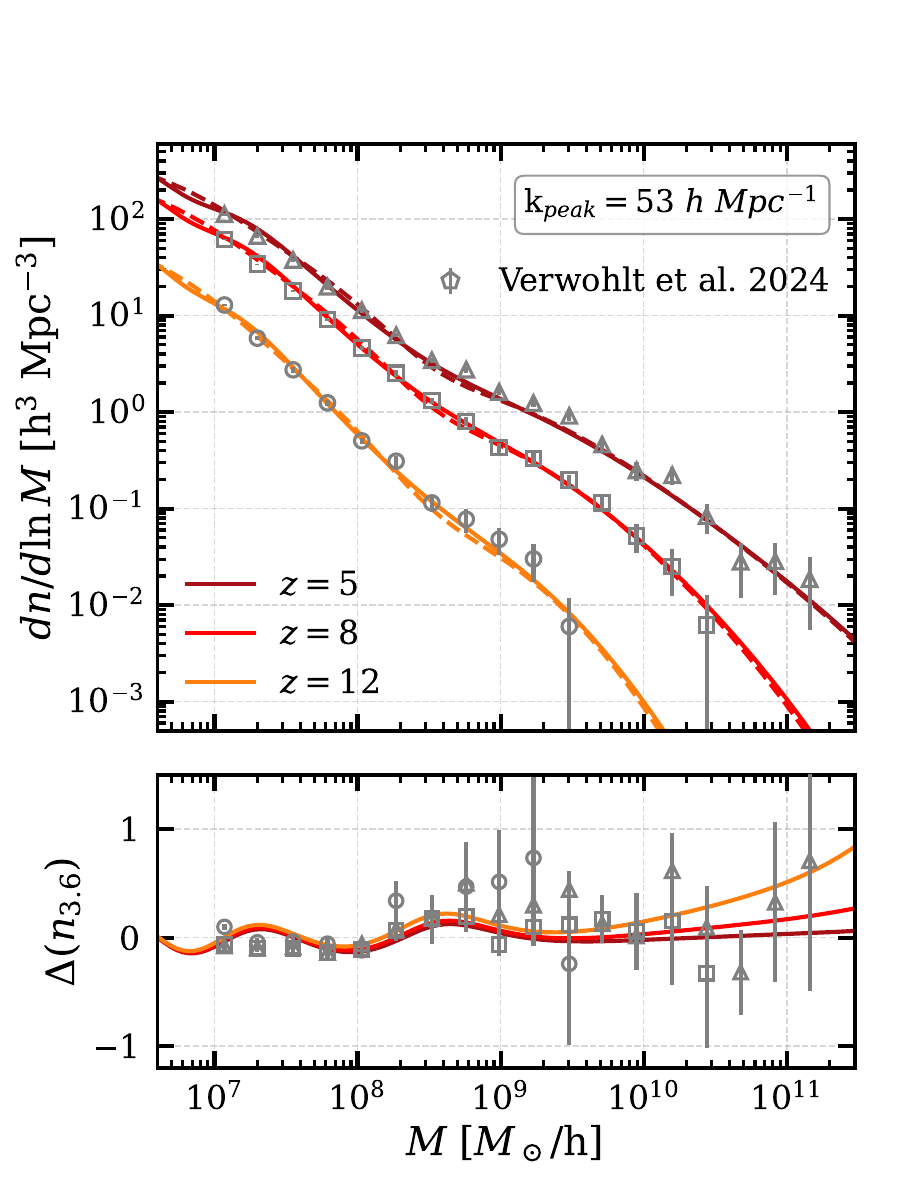}
    \end{minipage}
    \hfill
    \begin{minipage}{0.49\textwidth}
        \centering
        \includegraphics[width=\textwidth]{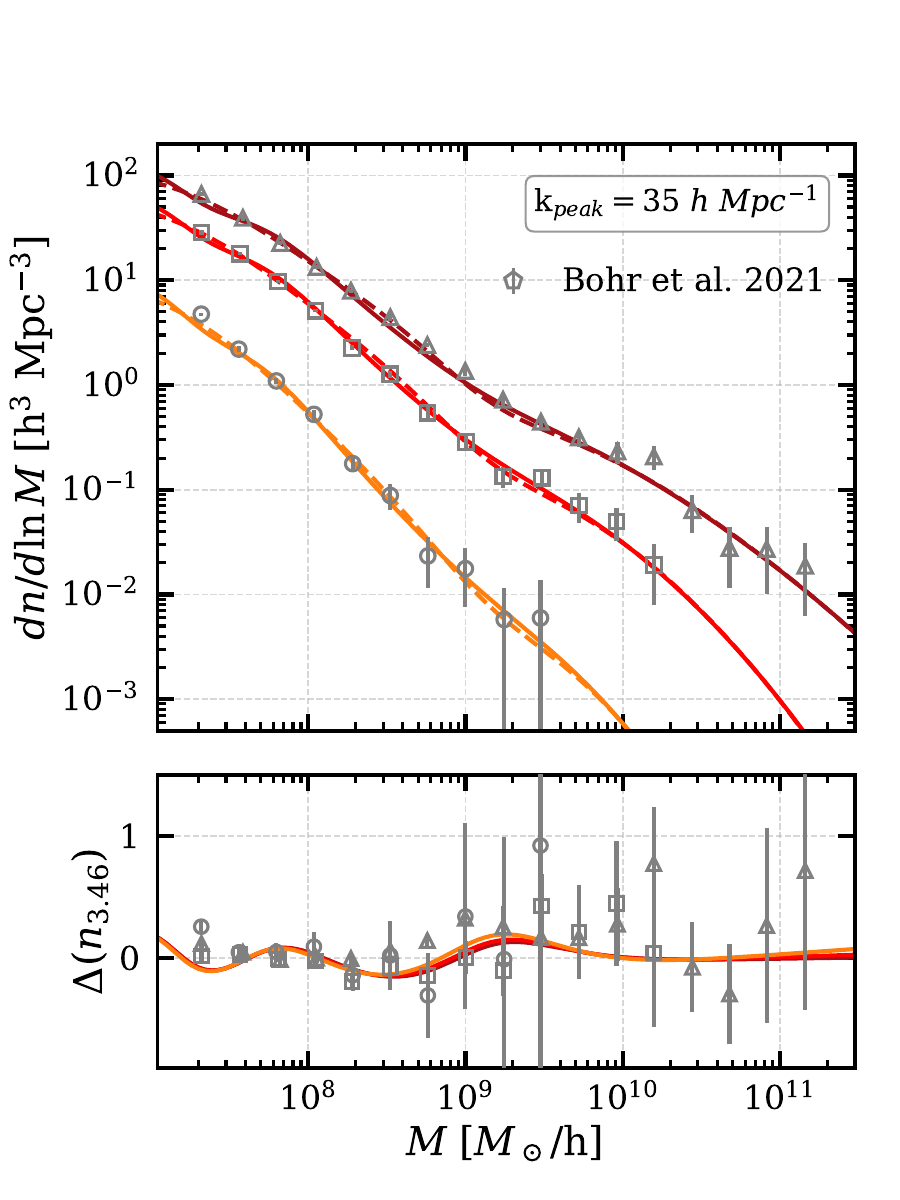}
    \end{minipage}
    \caption{\textbf{Consistency of the VSMK formalism with previously calibrated DAO halo mass functions.}
    Halo mass functions computed with the VSMK filter (solid lines) are compared with those obtained using the corresponding SMK calibrations reported in \cite{Verwohlt2024,Bohr2021} (dashed lines) for a range of DAO models and redshifts.
    \textit{Left:} DAO model from \cite{Verwohlt2024}, where the optimal SMK filter has $\beta = 3.6$ and $c = 3.6$.
    \textit{Right:} DAO model from \cite{Bohr2021}, where the optimal SMK filter has $\beta = 3.46$ and $c = 3.79$.
    Simulation data (empty circles) are extracted from published figures in \cite{Verwohlt2024} and \cite{Bohr2021}.
    The lower panels show the relative deviations $\Delta$ with respect to the optimal SMK prediction. The transfer-function parameter $h_{\mathrm{peak}}$ is fixed to unity, while $k_{\mathrm{peak}}$ is indicated in each panel.}
    \label{fig:results2}
\end{figure}

To evaluate the differences between halo mass functions obtained using different filters, we define the relative deviation between two halo mass functions as
\begin{equation}
\Delta_i(n_j) \equiv \Delta_i(M, n_j(M)) =
\frac{n_i(M) - n_j(M)}{n_j(M)},
\end{equation}
where $n_i(M)$ denotes the HMF computed with the filter under consideration, and $n_j(M)$ is the reference HMF. When the subscript is a numerical value, it corresponds to the slope $\beta$ of an SMK filter; when the subscript is VSMK, the VSMK filter with the parameters defined above is used. In the analyses presented here, $\Delta_i(n_j)$ is used both to quantify differences between analytical halo mass functions and to assess their agreement with $N$-body simulations. In the former case, it compares the VSMK prediction or a non-optimal SMK filter to the optimal SMK description of a given model (Figure~\ref{fig:results1}) or to previously reported SMK calibrations for DAO models (Figure~\ref{fig:results2}).

For the WDM model shown in Figure~\ref{fig:results1}, the deviation $\Delta_{2.8}(n_{4.8}) \to M^{-0.67} - 1$ as $M \rightarrow 0$, in agreement with Eq.~\eqref{eq:small_scale_HMF_trend}. In contrast, $\Delta_{\mathrm{VSMK}}(n_{4.8})$ remains below $0.25$ over the full mass range considered, reflecting the fact that the VSMK filter reproduces the correct small-scale asymptotic behavior. For the DAO model, the parameters $\mu = 2.1$ and $\delta = 12$ ensure that $\Delta_{\mathrm{VSMK}}(n_{2.8}) \leq 0.15$ within the resolved mass range, $10^7~M_\odot \lesssim M \lesssim 10^{10}~M_\odot$, whereas using the optimum SMK filter for the WDM model yields up to $\Delta_{4.8}(n_{2.8}) \approx 0.40$. 

As shown in Figure~\ref{fig:results1}, the VSMK filter reproduces the simulation data with a level of agreement comparable to that achieved by the optimal SMK calibration in each model. Table~\ref{tab:error} quantifies this result. For the WDM model, adopting the DAO-calibrated SMK filter ($\beta=2.8$) leads to substantially larger deviations from the simulation data, whereas the VSMK prescription remains comparable to the optimal WDM calibration ($\beta=4.8$, $c=3.3$). Likewise, for the DAO model, the VSMK filter achieves a level of agreement comparable to that obtained with the optimal SMK calibration while significantly improving upon the WDM-calibrated SMK filter. These results indicate that the VSMK formalism preserves the predictive power of model-specific SMK calibrations while providing a single parameterization capable of describing both WDM and DAO scenarios simulataneously.

\begin{table}[h]
\centering
\caption{Mean absolute relative deviations $\langle | \Delta_i(n_j) | \rangle$ between the analytic halo mass
functions $n_i$ and the $N$-body simulation data $n_j$ for the two cases shown in
Figure~\ref{fig:results1}. Values are reported as mean $\pm$ standard deviation of the
absolute relative deviations evaluated at the simulation data points. For both the WDM and DAO models, only data points with relative uncertainties below $25\%$ are considered. Unless stated otherwise, $c = 3.8$.}
\vspace{3mm}
\begin{tabular}{c c c c} 
\hline 
\noalign{\vskip 4pt} 
& ~ \shortstack{SMK (fit WDM) \\ ($\beta = 4.8$, $c = 3.3$)} 
& ~ ~ \shortstack{SMK (fit DAO) \\ ($\beta = 2.8$)} 
& ~ \shortstack{VSMK \\ ~ \\ ($\beta_1 = 4.8$, $\beta_2 = 2.8$)} \\ 
\noalign{\vskip 2pt} 
\hline 
\noalign{\vskip 4pt} 
WDM & ~ $0.07 \pm 0.09$ & ~ ~ $0.30 \pm 0.40$ & ~ $0.06 \pm 0.06$\\ 
DAO & ~ $0.18 \pm 0.13$ & ~ ~ $0.13 \pm 0.09$ & ~  $0.12 \pm 0.07$\\ 
\noalign{\vskip 2pt} 
\hline 
\end{tabular}
\label{tab:error}
\end{table}
\vspace{4mm}

Since our goal is to construct a generalized filter rather than a model-specific calibration, we further test the consistency of the VSMK formalism against a broader set of ETHOS simulations from \cite{Bohr2021,Verwohlt2024}. Figure~\ref{fig:results2} shows that the VSMK filter, characterized by $\beta_1 = 4.8$, $\beta_2 = 2.8$, $\mu = 2.1$, $\delta = 12$, and $c = 3.8$, reproduces the halo mass functions obtained across different ETHOS models with a level of agreement comparable to that achieved by previously reported SMK calibrations. Within the resolved mass range, the VSMK filter exhibits $\langle | \Delta_{\rm VSMK}(n_{\rm SMK}) | \rangle \lesssim 0.1$ for both reference SMK calibrations at all redshifts considered. Similarly, the agreement with the simulation measurements remains at the level
of $\langle | \Delta_{VSMK}(n_{data}) | \rangle \lesssim 0.12$ for all models and redshifts shown in Figure~\ref{fig:results2}, particularly in the mass ranges where the simulation uncertainties are below $25\%$.

These results show that a single VSMK parameterization remains consistent with the SMK descriptions previously reported for different ETHOS models and redshifts, while simultaneously providing a comparably good description of the underlying simulation data without requiring model-dependent recalibration.

\section{Conclusions}
\label{sec:conclusions}
Within the extended Press-Schechter framework, the smooth-$k$ filter has been shown to accurately describe the halo mass function for dark matter models featuring a small-scale suppression in the matter power spectrum. Its single free parameter is sufficient to reproduce the steep decline of the HMF at low halo masses, which is primarily determined by the large-scale behavior of the filter. This approach, however, becomes insufficient when the power spectrum exhibits multiple physically distinct regimes, such as a sharp suppression at very small scales combined with a damped oscillatory behavior at intermediate scales, as encountered in models with dark acoustic oscillations. In these cases, adjusting the small-mass slope of the HMF through a single SMK parameter inevitably induces an unwanted modification of the intermediate-mass regime, precisely where acoustic oscillations leave their imprint.

To address this limitation, we have introduced the variable-slope smooth-$k$ (VSMK) filter, which makes this scale dependence explicit by allowing for two distinct functional regimes characterized by independent values of the slope parameter. This construction enables a controlled association between the behavior of the filter at large spatial scales and the slope of the HMF at small masses, while the behavior of the filter at smaller scales independently regulates the intermediate-mass regime affected by DAO-induced oscillations. In this sense, the VSMK filter makes explicit a scale-inversion mechanism that, in the WDM case, remains hidden due to the simplicity of the underlying power spectrum. From a physical perspective, the variation of the effective slope parameter across mass regimes can be interpreted within the Press-Schechter framework as reflecting the different weights with which density perturbations at distinct scales contribute to the formation of low- and high-mass halos.

The VSMK filter thus constitutes a minimal functional extension of the SMK filter, introducing only the additional degrees of freedom required to interpolate between the relevant regimes. This minimality allows for a consistent treatment of damped and oscillatory power spectra, such as those associated with DAO models, without resorting to model-specific re-calibrations. By simultaneously reproducing halo mass functions from independent WDM and DAO simulation suites, we identify a common VSMK parameterization, characterized by $\beta_1 = 4.8$, $\beta_2 = 2.8$ and $c = 3.8$, that successfully reproduces the existing WDM and DAO calibrations within a single analytical framework. This provides a unified analytical description of structure formation in non-cold dark matter scenarios. Future observations and simulations probing progressively smaller halo masses will offer new opportunities to test the asymptotic predictions of the model and refine the description of halo abundances in the low-mass regime.

\appendix
    \section{Impact of the filter asymptotic behavior on different HMF mass regimes}
    \label{app:aprox_filter}
    In Section~\ref{sec:filter}, we showed that the position of the maximum of the integrand of the variance derivative,
    $\xi(k,k_M)$, provides direct insight into the asymptotic behavior of the HMF.
    This approach is particularly transparent in the small-mass regime for damped power spectra, where $\sigma^2(R)\to \text{constant}$ and the shape of the HMF is therefore primarly determined by $d \sigma^2(R) / d k_M$. Additionally, it sheds light on the behavior at intermediate scales for damped and oscillatory power spectra, as discussed below.
    
    \subsection{Filter behavior at $k/k_M \ll 1$ controls the small-mass limit of the HMF}
    \label{app:small_k_small_mass}
    Since the suppression of the HMF at low masses is driven by the small-scale damping of the matter
    power spectrum --- described either by the transfer function in Eq.~\eqref{eq:T_WDM} or by the leading
    term of Eq.~\eqref{eq:T_DAO} --- we consider a generic power spectrum featuring such a
    suppression. As a concrete example, the linear power spectrum of a WDM model with
    $m_{\mathrm{WDM}} = 0.25\,\mathrm{keV}$, obtained by applying the transfer function
    in Eq.~\eqref{eq:T_WDM} to the CDM power spectrum, exhibits the three characteristic asymptotic regimes of damped power spectra \cite{Coles1995,Leo2018}:
    \begin{equation}
        \label{eq:example_power_spectrum}
        P_{\mathrm{WDM}}(k) \propto
        \begin{cases}
            k^{n_s}, & k \lesssim 10^{-2}, \\
            k^{-3}, & 10^{-1} \lesssim k \lesssim 1, \\
            k^{-24+n_s}, & k \gtrsim 10^{1},
        \end{cases}
    \end{equation}
    where the small spatial scale behavior is obtained explicitly in \cite{Leo2018}.

    Although the precise slopes and transition scales vary among different ETHOS models, their qualitative behavior at large wavenumbers is similar. In both WDM and ETHOS models with dark acoustic oscillations (also referred to as DAO), the small-scale suppression is governed by a term with the same functional form as the leading contribution in Eq.~\eqref{eq:T_DAO}. Consequently, the arguments developed below apply generally to both WDM and DAO models, regardless of the specific particle mass or ETHOS parameters, with only quantitative differences.

\begin{figure}[htbp]
    \centering
    \begin{minipage}{0.47\textwidth}
        \centering
        \includegraphics[width=\textwidth]{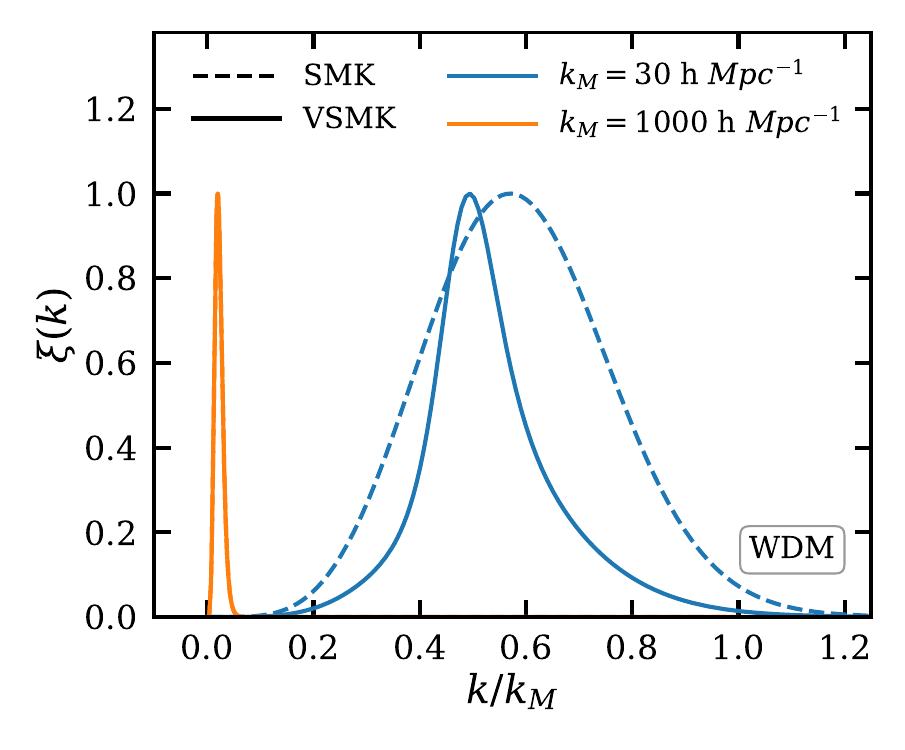}
    \end{minipage}
    \hspace{3mm}
    \begin{minipage}{0.47\textwidth}
        \centering
        \includegraphics[width=\textwidth]{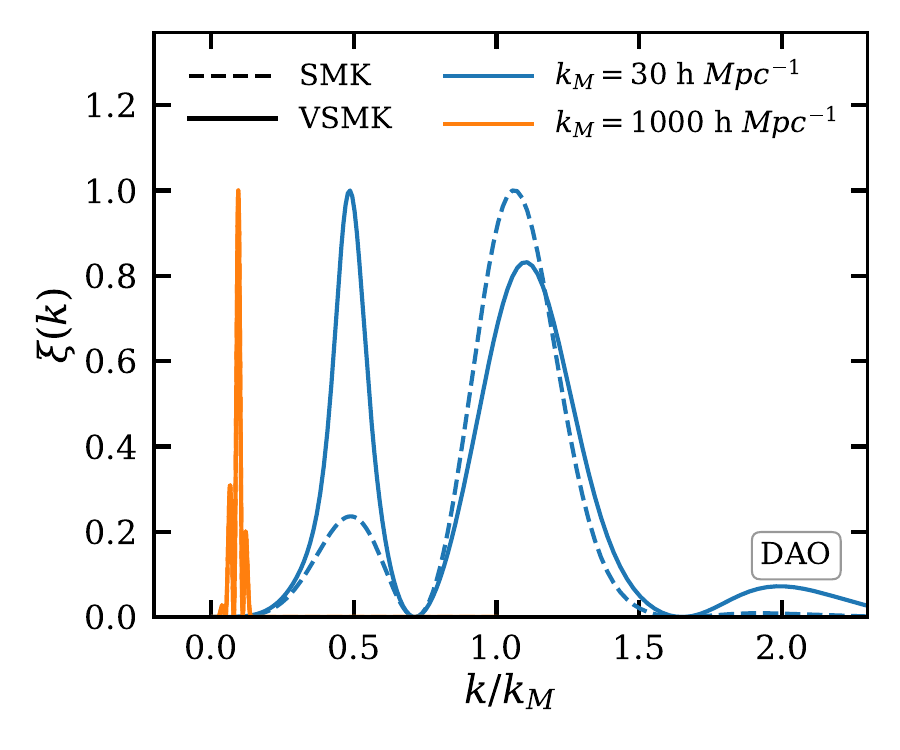}
    \end{minipage}
        \caption{\textbf{Normalized integrand of the variance derivative for WDM and DAO models.}
        The normalized integrands of the variance derivative corresponding to large and small mass scales are presented for the two dark matter models considered in this work. Both the SMK filter (dashed lines), with
        $\beta = 4.8$ and $c = 3.3$ and the VSMK filter (solid lines), with $\beta_1 = 4.8$, $\beta_2 = 2.8$, $\mu = 2.1$, $\delta = 12$ and $c = 3.8$ (Section~\ref{sec:results}) are shown for each model. \textit{Left:} WDM model with
        $m_{\mathrm{WDM}} = 1.61\,\mathrm{keV}$. \textit{Right:} DAO model with
        $k_{\mathrm{peak}} = 35\,h\,\mathrm{Mpc}^{-1}$ and $h_{\mathrm{peak}} = 1$.
        Each integrand is normalized to its own maximum.}
        \label{fig:integrand}
    \end{figure}

    The VSMK filter approaches a SMK filter with $\beta = \beta_1$ for $k/k_M < 1/\mu \sim O(1)$, and a SMK filter with $\beta = \beta_2$ for $k/k_M > 1/\mu$ as directly follows from Eqs.~\eqref{eq:VSMK} and \eqref{eq:variable_beta_VSMK}. Therefore, to analyze the position and shape of the maximum of \(\xi(k,k_M)\) \eqref{eq:integrand}, it is useful to consider the partial
    derivative of the SMK filter with respect to the inverse of the scale \(k_M = R^{-1}\), which enters
    explicitly in the integrand:
    \begin{equation}
        \frac{\partial W_{\mathrm{SMK}}(k,R)}{\partial k_M}
        =
        \frac{\beta \left( k / k_M \right)^{\beta}}
        {k_M \left[ 1 + \left( k / k_M \right)^{\beta} \right]^2}.
    \end{equation}
    This derivative attains its maximum at $k = k_M$, and increasing \(k_M\) shifts this peak toward progressively smaller
    physical scales in the power spectrum. When \(k_M\) becomes sufficiently large, $k_M \gg k_{\text{hm}}$, the maximum of the
    filter derivative lies within the small-scale regime of the power spectrum (e.g., \(k \gtrsim
    10\) in the WDM example above). As a result, the peak of the filter derivative is suppressed and the maximum of the integrand \(\xi(k,k_M)\) no longer occurs
    near \(k / k_M \simeq 1\), but instead moves to the regime \(k / k_M \ll 1\), closer to the maximum of the power spectrum itself (see Figure~\ref{fig:integrand}).

    For sufficiently steep filter slopes, specifically for $\beta > 22 - n_s$, such that the filter slope exceeds the absolute value of the slope of $k^2 P(k)$ implied by Eq.~\eqref{eq:T_WDM}, the decay of the
    power spectrum is no longer steep enough to supress the peak of the filter derivative. In this case,
    the maximum of \(\xi(k,k_M)\) remains located near \(k \simeq k_M\) and the resulting small-mass behavior of the HMF closely approaches that obtained
    with a sharp-$k$ filter, where the asymptotic regime is directly controlled by the power spectrum, $\mathrm{d}n/\mathrm{d}\ln M \propto k_M^6 P_{\mathrm{WDM}}(k_M)$ in the large-$k_M$ limit \cite{Schneider2013}. This behavior can be straightforwardly derived from Eqs.~\eqref{eq:deriv_var} and \eqref{eq:integrand}, noting that the derivative of the SHK filter with respect to $k_M$ reduces to a Dirac delta.

    \subsection{Independence of the intermediate-mass HMF from the filter in WDM models}
    \label{sec:appendix_WDM_filter_independence}

    At intermediate mass scales, the variance $\sigma^2(R)$ defined in Eq.~\eqref{eq:variance} decreases monotonically as the scale $R = k_M^{-1}$ increases \cite{Bose2016}. In this regime, however, the first-crossing distribution given by Eq.~\eqref{eq:first_crossing_distribution} varies only slowly with $\sigma^2(R)$ \cite{Xu2023}. The arguments presented in the previous subsection therefore make it possible to identify the dominant contribution to the slope of the HMF.

    Specifically, provided that the filter slope satisfies $\beta > 1$, 
    the maximum of the integrand $\xi(k,k_M)$ defined in Eq.~\eqref{eq:integrand} lies outside 
    the steeply suppressed small-scale regime and instead falls within the second asymptotic regime of the linear matter power spectrum. For sufficiently small values of $\beta$, the absolute value of the slope of $k^2 P(k)$ overcomes $\beta$ and the maximum of $\xi(k,k_M)$ is dominated by the maximum of the power spectrum itself.

    In the intermediate- and large-mass regimes, varying the halo mass $M$ primarily shifts the peak of $W\,\partial W/\partial k_M$ along a portion of the power spectrum characterized by a smooth power-law behavior with an approximately constant slope, corresponding to the second regime of the example power spectrum in Eq.~\eqref{eq:example_power_spectrum}. As a result, the integral determining $d\sigma^2/dk_M$ consistently probes the region $k/k_M \lesssim 1$ (see Figure~\ref{fig:integrand}), where the integrand varies with an approximately constant slope, largely independent of the precise value of $\beta$. Combined with the weak dependence of the first-crossing distribution on $\sigma^2$ at intermediate and large masses ($10^8$--$10^{14}\,M_\odot$) \cite{Xu2023}, this implies that the halo mass function in this range is only mildly sensitive to the choice of filter, as shown in Figure~\ref{fig:HMF}.

    In the limiting case $\beta \to \infty$, the smooth-$k$ filter reduces to the sharp-$k$ filter. In this limit, the derivative of the filter becomes a Dirac delta function centered at $k = k_M$, so that the derivative of the variance can be evaluated analytically and depends solely on the value of the power spectrum at $k_M$. This further illustrates that, for WDM models, the intermediate- and large-mass behavior of the HMF is controlled primarily by the power spectrum rather than by the detailed shape of the filter.

    \subsection{Filter dependence of the intermediate-mass HMF in models with oscillatory power spectra}
    \label{sec:appendix_DAO_filter_dependence}

    The situation changes qualitatively in models with dark acoustic oscillations, where the linear matter power spectrum exhibits damped oscillatory features at intermediate scales. In this case, the dependence of the halo mass function on the filter can be qualitatively understood by extending the arguments presented above.

    For a WDM model with $\beta > 1$, the integrand $\xi(k,k_M)$ displays a single dominant peak located near $k \simeq k_M$ at sufficiently large halo masses. However, in models with oscillatory power spectra, as $k_M$ increases and intermediate masses are probed, this peak may no longer lie in a region where the power spectrum is smooth. Instead, it coincides with the DAO-induced oscillations, as shown in Figure~\ref{fig:integrand}. Depending on the relative phase, the peak of $\xi(k,k_M)$ may be enhanced or suppressed, in close analogy with the small-mass regime of WDM or DAO models, where the peak is diminished once it enters the steeply damped region of the power spectrum.

    As a consequence, oscillatory power spectra can give rise to multiple local maxima in $\xi(k,k_M)$. This occurs because the maximum of the filter derivative can remain significant even beyond the first oscillation, since the power spectrum does not decay monotonically but instead partially recovers its amplitude after each oscillation \cite{Schaeffer2021,Verwohlt2024}. In this regime, the power spectrum alone no longer uniquely determines the dominant contribution to $d\sigma^2/dk_M$; the detailed shape of the filter becomes equally important.

    The role of the filter is to regulate the relative weight of contributions away from the primary peak that would arise in a WDM-like scenario. In particular, the filter smooths the oscillatory features of the power spectrum, thereby controlling the amplitude and persistence of the corresponding oscillations in the halo mass function. In the limiting case $\beta \to \infty$, corresponding again to the SHK filter, the derivative of the filter reduces to a Dirac delta function. As in the WDM case, the derivative of the variance becomes directly proportional to the power spectrum, leading to a highly nonphysical halo mass function that vanishes at masses corresponding to minima of the oscillatory power spectrum \cite{Schaeffer2021}.

\acknowledgments
The author is especially grateful to Claudi Vall for sustained and in-depth discussions throughout the various stages of this project, which were instrumental in shaping the ideas presented here. The author also acknowledges valuable discussions with Alberto Manrique, Chervin Laporte, Jordi Martorell and Judith López Zancajo. This research received no external funding.

\paragraph{Data and Code availability.}
No new data were generated or analysed in this study. The numerical routines used to evaluate the analytical expressions presented in this work are available from the author upon reasonable request.


\end{document}